\documentclass[aps,prd,superscriptaddress,preprint,tightenlines,nofootinbib,showpacs]{revtex4-1}

\usepackage{graphicx} 
\usepackage{dcolumn}  
\usepackage{bm}       

\newcommand{\Dzkpi}{\Dz\to\Km\pip}
\newcommand{\Dzkpipiz}{\Dz\to\Km\pip\piz}
\newcommand{\Dzkpipipi}{\Dz\to\Km\pip\pip\pim}
\newcommand{\Dpkpipi}{\Dp\to\Km\pip\pip}
\newcommand{\Dpkpipipiz}{\Dp\to\Km\pip\pip\piz}
\newcommand{\Dpkspi}{\Dp\to\KS\,\pip}
\newcommand{\Dpkspipiz}{\Dp\to\KS\,\pip\piz}
\newcommand{\Dpkspipipi}{\Dp\to\KS\,\pip\pip\pim}
\newcommand{\Dpkkpi}{\Dp\to\Kp\Km\pip}

\newcommand{\Dzbarkpi}{\Dzbar\to\Kp\pim}
\newcommand{\Dzbarkpipiz}{\Dzbar\to\Kp\pim\piz}
\newcommand{\Dzbarkpipipi}{\Dzbar\to\Kp\pim\pim\pip}
\newcommand{\Dmkpipi}{\Dm\to\Kp\pim\pim}
\newcommand{\Dmkpipipiz}{\Dm\to\Kp\pim\pim\piz}
\newcommand{\Dmkspi}{\Dm\to\KS\,\pim}
\newcommand{\Dmkspipiz}{\Dm\to\KS\,\pim\piz}
\newcommand{\Dmkspipipi}{\Dm\to\KS\,\pim\pim\pip}
\newcommand{\Dmkkpi}{\Dm\to\Km\Kp\pim}

\newcommand{\epemDzDzbar}{\elp\elm\to\Dz\Dzbar}
\newcommand{\epemDpDm}{\elp\elm\to\Dp\Dm}

\newcommand{\cleoc}{\hbox{CLEO-c}}

\newcommand{\elp}{e^+}
\newcommand{\elm}{e^-}

\newcommand{\pip}{\pi^+}
\newcommand{\pim}{\pi^-}
\newcommand{\piz}{\pi^0}
\newcommand{\pipm}{\pi^\pm}

\newcommand{\Kp}{K^+}
\newcommand{\Km}{K^-}

\newcommand{\KS}{K^0_S}

\newcommand{\Kpm}{K^\pm}

\newcommand{\Dp}{D^+}
\newcommand{\Dm}{D^-}
\newcommand{\Dz}{D^0}
\newcommand{\Dbar}{\overline{D}}
\newcommand{\Dzbar}{\overline{D}{}^0}

\newcommand{\BDzkpivalue}{3.934 \pm 0.021 \pm 0.061}
\newcommand{\BDpkpipivalue}{9.224 \pm 0.059 \pm 0.157}
\newcommand{\sigDzDzbarvalue}{3.607\pm 0.017 \pm 0.056}
\newcommand{\sigDpDmvalue}{2.882\pm 0.018 \pm 0.042}
\newcommand{\sigDDbarvalue}{6.489\pm 0.024 \pm 0.092}
\newcommand{\sigDDbarratio}{0.799\pm 0.006 \pm 0.008}

\newcommand{\BDzkpi}{\calB(\Dzkpi)}

\newcommand{\BDpkpipi}{\calB(\Dpkpipi)}

\newcommand{\Ecm}{E_\mathrm{cm}}

\newcommand{\Egamma}{E_\gamma}

\newcommand{\Gev}{\mathrm{GeV}}
\newcommand{\Gevc}{\mathrm{GeV}/c}

\newcommand{\Mev}{\mathrm{MeV}}

\newcommand{\DeltaE}{\Delta E}

\newcommand{\Mbc}{M_\mathrm{BC}}

\newcommand{\bfp}{\mathbf{p}}

\newcommand{\calB}{\mathcal{B}}

\newcommand{\pbinv}{pb$^{-1}$}

\newcommand{\jbar}{\bar{\jmath}}

\newcommand{\Bi}{\calB_i}
\newcommand{\Bj}{\calB_j}
\newcommand{\Ni}{y_i}

\newcommand{\effi}{\epsilon_i}
\newcommand{\effj}{\epsilon_j}
\newcommand{\Bjbar}{\calB_{\jbar}}

\newcommand{\Njbar}{y_{\jbar}}

\newcommand{\effjbar}{\epsilon_{\jbar}}

\newcommand{\Nijbar}{y_{i\jbar}}

\newcommand{\effijbar}{\epsilon_{i\jbar}}

\newcommand{\NDDbar}{N_{D\Dbar}}
\newcommand{\NDzDzbar}{N_{\Dz\Dzbar}}
\newcommand{\NDpDm}{N_{\Dp\Dm}}

\newcommand{\nf}{n(f)}
\newcommand{\nfbar}{n(\overline{f})}

\newcommand{\ie}{\textit{i.e.}}

\newlength{\Plotwidth}
\newcommand{\Begitem}{\begin{itemize}}
\newcommand{\Enditem}{\end{itemize}}

\newcommand{\Tab}[1]{Table~\ref{#1}}
\newcommand{\Ref}[1]{Ref.~\cite{#1}}

\newcommand{\Begeqn}{\begin{equation}}
\newcommand{\Endeqn}{  \end{equation}}

\begin{document}

\preprint{CLNS 13/2087}  
\preprint{CLEO 13-02}    

\title{\boldmath Updated measurements of absolute $D^+$ and $\Dz$ hadronic branching fractions and $\sigma(\elp\elm \to D\Dbar)$ at $\Ecm = 3774$ MeV}

\author{G.~Bonvicini}
\author{D.~Cinabro}
\author{M.~J.~Smith}
\author{P.~Zhou}
\affiliation{Wayne State University, Detroit, Michigan 48202, USA}
\author{P.~Naik}
\author{J.~Rademacker}
\affiliation{University of Bristol, Bristol BS8 1TL, United Kingdom}
\author{K.~W.~Edwards}
\affiliation{Carleton University, Ottawa, Ontario K1S 5B6, Canada}
\author{R.~A.~Briere}
\author{H.~Vogel}
\affiliation{Carnegie Mellon University, Pittsburgh, Pennsylvania 15213, USA}
\author{J.~L.~Rosner}
\affiliation{University of Chicago, Chicago, Illinois 60637, USA}
\author{J.~P.~Alexander}
\author{D.~G.~Cassel}
\author{R.~Ehrlich}
\author{L.~Gibbons}
\author{S.~W.~Gray}
\author{D.~L.~Hartill}
\author{B.~K.~Heltsley}
\author{D.~L.~Kreinick}
\author{V.~E.~Kuznetsov}
\author{J.~R.~Patterson}
\author{D.~Peterson}
\author{D.~Riley}
\author{A.~Ryd}
\author{A.~J.~Sadoff}
\author{X.~Shi}
\altaffiliation[Present address: ]{Purdue University, West Lafayette, Indiana 47097, USA.}
\author{W.~M.~Sun}
\affiliation{Cornell University, Ithaca, New York 14853, USA}
\author{S.~Das}
\author{J.~Yelton}
\affiliation{University of Florida, Gainesville, Florida 32611, USA}
\author{P.~Rubin}
\affiliation{George Mason University, Fairfax, Virginia 22030, USA}
\author{N.~Lowrey}
\author{S.~Mehrabyan}
\author{M.~Selen}
\author{J.~Wiss}
\affiliation{University of Illinois, Urbana-Champaign, Illinois 61801, USA}
\author{J.~Libby}
\affiliation{Indian Institute of Technology Madras, Chennai, Tamil Nadu 600036, India}
\author{M.~Kornicer}
\author{R.~E.~Mitchell}
\affiliation{Indiana University, Bloomington, Indiana 47405, USA }
\author{D.~Besson}
\affiliation{University of Kansas, Lawrence, Kansas 66045, USA}
\author{T.~K.~Pedlar}
\affiliation{Luther College, Decorah, Iowa 52101, USA}
\author{D.~Cronin-Hennessy}
\author{J.~Hietala}
\affiliation{University of Minnesota, Minneapolis, Minnesota 55455, USA}
\author{S.~Dobbs}
\author{Z.~Metreveli}
\author{K.~K.~Seth}
\author{A.~Tomaradze}
\author{T.~Xiao}
\affiliation{Northwestern University, Evanston, Illinois 60208, USA}
\author{A.~Powell}
\author{C.~Thomas}
\author{G.~Wilkinson}
\affiliation{University of Oxford, Oxford OX1 3RH, United Kingdom}
\author{D.~M.~Asner}
\author{G.~Tatishvili}
\affiliation{Pacific Northwest National Laboratory, Richland, Washington 99352, USA}
\author{J.~Y.~Ge}
\author{D.~H.~Miller}
\author{I.~P.~J.~Shipsey}
\altaffiliation[Present address: ]{University of Oxford, Oxford OX1 3RH, United Kingdom.}
\author{B.~Xin}
\affiliation{Purdue University, West Lafayette, Indiana 47907, USA}
\author{G.~S.~Adams}
\author{J.~Napolitano}
\affiliation{Rensselaer Polytechnic Institute, Troy, New York 12180, USA}
\author{K.~M.~Ecklund}
\affiliation{Rice University, Houston, Texas 77005, USA}
\author{J.~Insler}
\author{H.~Muramatsu}
\author{L.~J.~Pearson}
\author{E.~H.~Thorndike}
\affiliation{University of Rochester, Rochester, New York 14627, USA}
\author{M.~Artuso}
\author{S.~Blusk}
\author{R.~Mountain}
\author{T.~Skwarnicki}
\author{S.~Stone}
\author{J.~C.~Wang}
\author{L.~M.~Zhang}
\affiliation{Syracuse University, Syracuse, New York 13244, USA}
\author{P.~U.~E.~Onyisi}
\affiliation{University of Texas at Austin, Austin, Texas 78712, USA}
\collaboration{CLEO Collaboration}
\noaffiliation

\date{\today}

\begin{abstract}
  
Utilizing the full CLEO-c data sample of 818~\pbinv\ of $\elp\elm$ data 
taken at the $\psi(3770)$ resonance, we update our measurements of absolute hadronic 
branching fractions of charged and neutral $D$ mesons. We previously  
reported results from subsets of these data.
  Using a double tag technique we obtain branching
  fractions for three $\Dz$ and six $\Dp$ modes, including the
  reference branching fractions ${\calB}(\Dzkpi)=(\BDzkpivalue)\%$ and
  ${\calB}(\Dpkpipi)=(\BDpkpipivalue)\%$.  The uncertainties are
  statistical and systematic, respectively.  In these measurements we
  include the effects of final-state radiation by allowing for
  additional unobserved photons in the final state, and the systematic
  errors include our estimates of the uncertainties of these effects.
  Furthermore, using an independent measurement of the luminosity, we
  obtain the cross sections $\sigma(\epemDzDzbar)=(\sigDzDzbarvalue) \
  \mathrm{nb}$ and $\sigma(\epemDpDm)=(\sigDpDmvalue) \ \mathrm{nb}$
  at a center of mass energy, $\Ecm = 3774 \pm 1$ MeV.
\end{abstract}


\maketitle


\section*{Introduction}

Precision measurements of absolute hadronic $D$ meson branching
fractions are essential for both charm and beauty physics.  For
example, determination of the Cabibbo-Kobayashi-Maskawa (CKM)
\cite{Cabibbo,Kobayashi} matrix element $|V_{cb}|$ utilizing the
exclusive decay $B \to D^{*} \ell \nu$ with full $D^*$ reconstruction
requires knowledge of the absolute $D$ meson branching fractions
\cite{pdg2012}. We report absolute measurements of three $\Dz$ and six
$\Dp$ branching fractions (averaged between $\Dz$ and $\Dzbar$ or
$\Dp$ and $\Dm$) for the Cabibbo-favored decays $\Dzkpi$,
$\Dzkpipiz$, $\Dzkpipipi$, $\Dpkpipi$, $\Dpkpipipiz$, $\Dpkspi$,
$\Dpkspipiz$, $\Dpkspipipi$, and the Cabibbo-suppressed decay
$\Dpkkpi$. We call $\BDzkpi$ and $\BDpkpipi$ reference branching
fractions because most $\Dz$ and $\Dp$ branching fractions are
determined from ratios to one of these branching fractions
\cite{pdg2012}.  

The data sample was produced in $e^+e^-$ collisions at the Cornell
Electron Storage Ring (CESR) and collected with the CLEO-c detector
\cite{cleoiidetector,cleoiiidr,cleorich,cleocyb}.  
It consists of 818~\pbinv\ of integrated luminosity collected 
on the $\psi(3770)$ resonance, at a center-of-mass energy 
$\Ecm=3774 \pm 1$ MeV. We previously
reported results based on 56~\pbinv\ \cite{cleodhadprl} and
281~\pbinv\ \cite{cleodhadprd281} subsamples of these data. 
These final measurements from CLEO supersede the earlier CLEO results.
Because the principal analysis technique is unchanged and was 
documented in great detail in \Ref{cleodhadprd281}, we will briefly
review the procedure here and focus primarily on significant improvements.

In accord with our previous measurements
\cite{cleodhadprl,cleodhadprd281}, we employ a ``double tagging''
technique pioneered by the MARK III Collaboration \cite{markiii-1,
  markiii-2} to measure these branching fractions. This technique
takes advantage of a unique feature of data taken at a center-of-mass
energy near the peak of the $\psi$(3770) resonance in $\elp\elm$
collisions.  This resonance is just above the threshold for $D\Dbar$ 
production, so only $\Dz\Dzbar$ and $\Dp\Dm$ pairs are produced without
additional hadrons in the final states.  We select ``single tag'' (ST)
events in which either a $D$ or $\Dbar$ is reconstructed without
reference to the other particle and ``double tag'' (DT) events in
which both the $D$ and $\Dbar$ are reconstructed.  Then we determine
absolute branching fractions for $\Dz$ or $\Dp$ decays from the
fraction of DT events in our ST samples.
    
 Letting $\NDDbar$ be the number of $D\Dbar$ events (either
 $\Dz\Dzbar$ or $\Dp\Dm$) produced in the experiment, the observed
 yields, $\Ni$ and $\Njbar$, of reconstructed $D\to i$ and
 $\Dbar\to\jbar$ ST events will be
\begin{equation}
\label{eq:st}
\Ni = \NDDbar\, \Bi\, \effi ~~\mathrm{and}~~ 
\Njbar = \NDDbar\, \Bj\, \effjbar, 
\end{equation}
where $\Bi$ and $\Bj$ are branching fractions for $D \to i$ and $D \to
j$, with the assumption that charge-conjugation parity ($CP$)
violation is negligible so that $\Bj = \Bjbar$. However, the
efficiencies $\effj$ and $\effjbar$ for detection of these modes may
not be the same due to the charge dependencies of cross sections for
the scattering of pions and kaons on the nuclei of the detector
material.  Furthermore, the DT yield for $D\to i$ (signal mode) and
$\Dbar\to\jbar$ (tagging mode) will be
\begin{equation}
\label{eq:dt}
\Nijbar = \NDDbar\, \Bi\, \Bj\, \effijbar,
\end{equation}
 where $\effijbar$ is the efficiency for detecting double tag events
in modes $i$ and $\jbar$.  A combination of Eqs.\ (\ref{eq:st}) and
(\ref{eq:dt}) yields an absolute measurement of the branching fraction $\Bi$,
\begin{equation}
\Bi = {\Nijbar \over \Njbar}{\effjbar \over \effijbar}.
\end{equation}
Note that $\effijbar \approx \effi\,\effjbar$, so $\effijbar/\effjbar
\approx \effi$, and the measured value of $\Bi$ is quite insensitive to the value of $\effjbar$.

We utilize a least-squares technique to extract branching fractions
and $\NDDbar$ by combining ST and DT yields.  Although the $\Dz$ and
$\Dp$ yields are statistically independent, systematic effects and
misreconstruction resulting in cross feeds among the decay modes introduce 
correlations among their uncertainties.  Therefore, we fit $\Dz$ and $\Dp$ parameters simultaneously by minimizing a $\chi^2$ that includes statistical 
and systematic uncertainties and their correlations for all experimental
inputs \cite{brfit}.  In the fit, we include the ST and DT efficiencies 
and -- as described below -- correct the ST and DT yields for backgrounds 
that peak in the regions of the signal peaks.

\section*{Detector and Reconstruction}

We reconstruct charged tracks in the \cleoc\ detector using the
47-layer drift chamber \cite{cleoiiidr} and the coaxial 6-layer vertex
drift chamber \cite{cleocyb}.  For tracks that traverse all layers of
the drift chamber, the root-mean-square (rms) momentum resolution is
approximately 0.6\% at $p = 1~\Gevc$.  We detect photons in an
electromagnetic calorimeter containing about 7800 CsI(Tl) crystals
\cite{cleoiidetector}, whose rms photon energy resolution is 2.2\% at
$\Egamma = 1~\Gev$, and 5\% at $\Egamma = 100~\Mev$.  The solid angle
for detection of charged tracks and photons is 93\% of $4\pi$.
Particle identification (PID) information to separate $\Kpm$
from $\pipm$ is provided by measurements of ionization ($dE/dx$) in the central drift chamber \cite{cleoiiidr} and by a
cylindrical ring-imaging Cherenkov (RICH) detector \cite{cleorich}.  
Below about $p=0.7~\Gevc$, separation using only
$dE/dx$ is very effective and we utilize this technique alone.  Above
that momentum, we combine information from $dE/dx$ and the RICH
detector when both are available.  The solid angle of the RICH
detector is about 86\% of the solid angle of the tracking system,
leading to a modest decrease in PID effectiveness above $p=0.7~\Gevc$.
We describe the PID techniques and performance in more detail in
\Ref{cleodhadprd281}.  We reconstruct $\KS$ in the decay mode
$\KS\to\pip\pim$, without requiring PID for the charged pions.  

We study the response of the \cleoc\ detector
utilizing a GEANT-based \cite{geant} Monte Carlo (MC) simulation of
particle detection.  We use EVTGEN \cite{evtgen} to generate $D$ and
$\Dbar$ daughters and PHOTOS \cite{photos} to simulate final-state
radiation (FSR).
 
We identify $D$ meson candidates by their beam-constrained masses
($\Mbc$) and total energies. For each candidate, we calculate $\Mbc$
by substituting the beam energy, $E_0$, for the measured $D$ candidate
energy, \ie, $\Mbc\,c^2 \equiv (E_0^2 - \bfp_D^2c^2)^\frac{1}{2}$, 
where $\bfp_D$ is the momentum of
the $D$ candidate.  The beam-constrained mass has a rms resolution of about 2
MeV/$c^2$, which is dominated by the beam energy spread. For the total
energy selection, we define $\Delta E\equiv E_D - E_0$, where $E_D$ is
the sum of the $D$ candidate daughter energies. For further analysis,
we select $D$ candidates with $\Mbc$ greater than 1.83 GeV/$c^2$ and
$|\Delta E|$ within mode-dependent limits that are approximately $\pm
3\sigma$ \cite{cleodhadprd281}. For both ST and DT modes, we accept at
most one candidate per mode per event, where conjugate modes are
treated as distinct.  For ST candidates, we chose the candidate with
the smallest $\Delta E$, while for DT candidates, we take the
candidate whose average of $D$ and $\Dbar$ $\Mbc$ values, denoted by
$\widehat{M}$, is closest to the known $D$ mass.

\section*{Single Tag and Double Tag Yields}

We extract ST and DT yields from $\Mbc$ distributions in the samples
described above. We perform unbinned maximum likelihood fits in one
and two dimensions for ST and DT modes, respectively, to a signal
shape and one or more background components. The signal shape includes
the effects of beam energy smearing, initial-state radiation, the line
shape of the $\psi(3770)$, and reconstruction resolution.  The
background in ST modes is described by an ARGUS function
\cite{argusf}, which models combinatorial contributions. In DT modes,
backgrounds can be uncorrelated, where either the $D$ or $\Dbar$ is
misreconstructed, or correlated, where all the final state particles
in the event are correctly reconstructed but are mispartitioned among
the $D$ and $\Dbar$.  In fitting the two-dimensional $\Mbc( D)$
versus $\Mbc( \Dbar)$ distribution, we model the uncorrelated
background by a pair of functions, where one dimension is an ARGUS
function and the other is the signal shape.  We model the correlated
background by an ARGUS function in $\widehat{M}$ and a Gaussian in the
orthogonal variable, which is $[\Mbc( \Dbar)-\Mbc( D)]/2$.  In
\Ref{cleodhadprd281} we describe in detail the fit functions that we
use and the parameters that determine these functions.

\begin{figure*}[htb]
\includegraphics*[width=\linewidth]{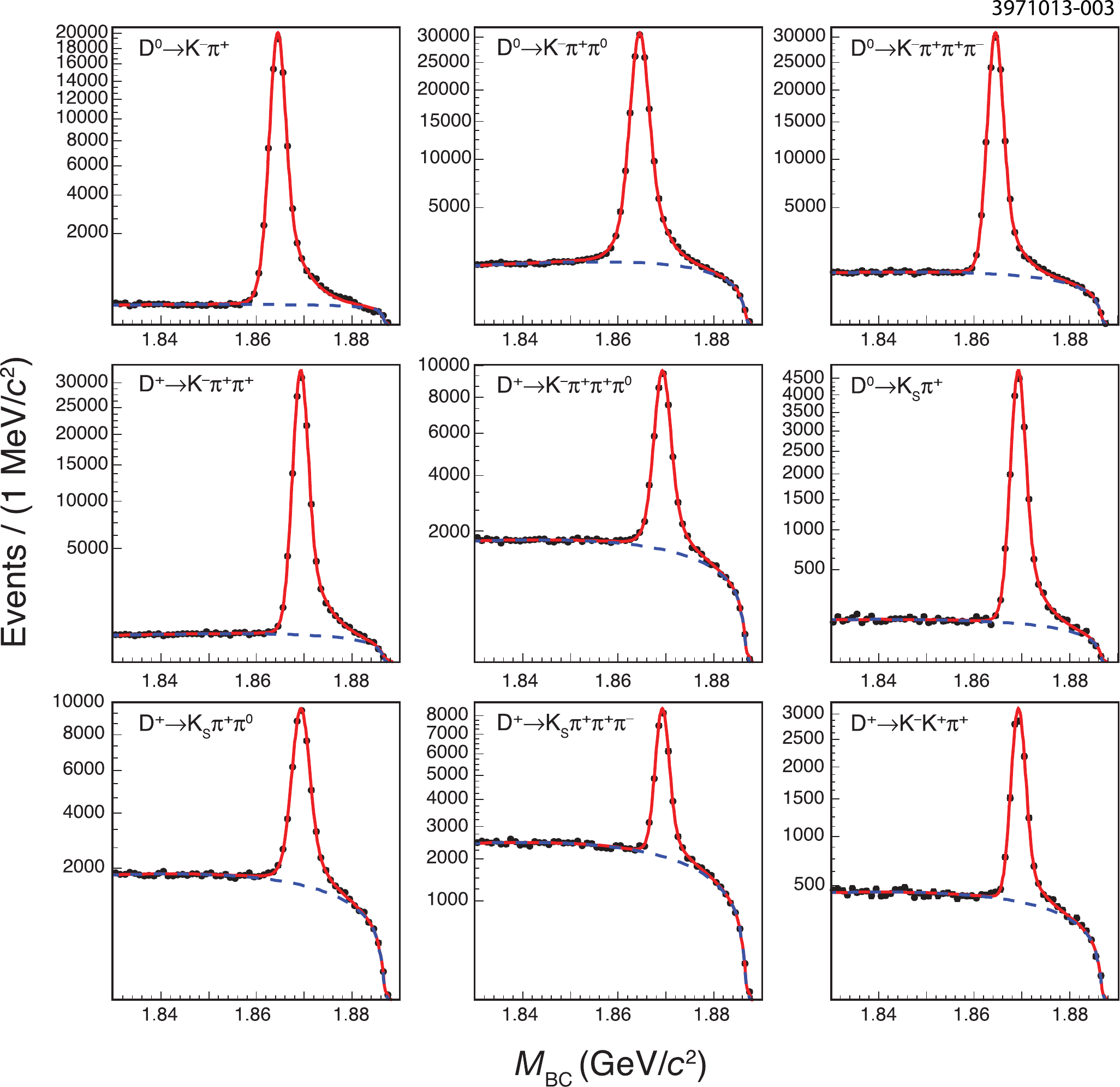}
\caption{Numbers of single tag event candidates, plotted on square-root scales,
  versus $\Mbc$ for each charged and neutral mode.  In each plot, $D$ and
  $\Dbar$ candidates are combined.  Data are shown as points and the solid lines (red online) show the total fits and the
  dashed lines (blue online) are the background shapes. The high-mass 
  tails on the signal are due to initial-state radiation. 
  \label{fig:data_single_tag}}
\end{figure*}

\begin{figure*}
\includegraphics*[width=\linewidth]{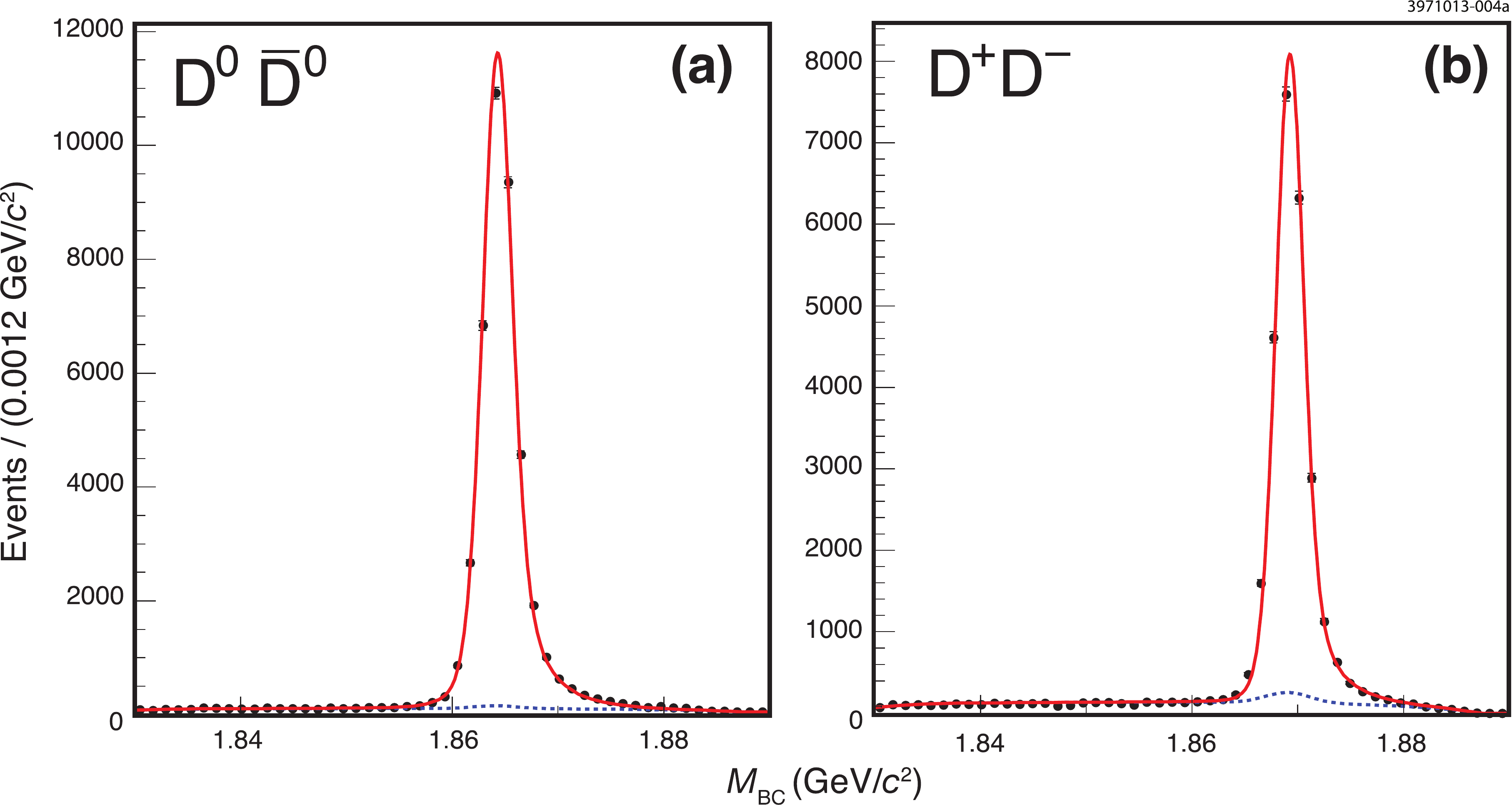}
\caption{Projections of double tag candidate  masses on the $\Mbc(D)$
  axis for (a) all $\Dz\Dzbar$ modes and (b) all $\Dp \Dm$ modes. In each
  plot, the points are data, the lines are projections of the fit results; 
  the dashed line 
  (blue online) is the peaking background contribution, and the solid line 
  (red online) is the sum of signal and background.
  \label{fig:data_double_tag}}
\end{figure*}

Table \ref{tab:st-eff-yield} gives the 18 ST data yields (without efficiency correction) and the corresponding
efficiencies, which are determined from simulated events.  
Figure~\ref{fig:data_single_tag} shows the $\Mbc$ 
distributions\footnote{We utilize square-root scales in 
Fig.~\ref{fig:data_single_tag}  
because these scales are an excellent visual compromise between   
linear scales (which emphasize signals) and logarithmic scales (which emphasize backgrounds).  This property results from the fact all error bars that are proportional to $\sqrt{N}$ are the same size on a square-root scale.  However, the error bars in these plots for small numbers of events are somewhat larger than the others because we utilize ROOFIT~\protect\cite{roofit} to produce these plots and ROOFIT error bars are 68\% confidence intervals.}
for the nine
decay modes with $D$ and $\Dbar$ candidates combined.  The fitted signal and background components are overlaid.  We also measure 45 DT
yields in data and determine the corresponding efficiencies from
simulated events.  Figure~\ref{fig:data_double_tag} shows projections
on the $\Mbc( D)$ axis for all (a) $\Dz\Dzbar$ and (b) $\Dp\Dm$ DT 
candidates.

Backgrounds with smooth $\Mbc$ distributions are well represented by 
ARGUS functions and do not contribute to the ST and DT yields, but 
there are backgrounds that peak in the signal regions
that do contribute to these yields.  
In the branching fraction fit, we correct the ST and DT yields for
two types of peaking backgrounds, which we call ``internal'' and 
``external''.  Internal or cross feed backgrounds come from decays to  
any one of our signal modes, $i$, that peak in the $\Mbc$ distributions 
of any other modes due to misreconstruction.  This type of contribution 
to any signal mode is proportional to the branching fraction $\Bi$ 
for the misreconstructed decay mode and the appropriate $\NDDbar$, 
both of which are determined in the fit.     
On the other hand, external backgrounds are from $D$ or $\Dbar$ decays to  
modes that we do not measure in this analysis, but which appear in the peaks of
signal modes due to misreconstruction.  These contributions are proportional 
to the appropriate $\NDDbar$ values that we obtain in the fit, and the branching fractions for the modes that we obtain from the particle data group \cite{pdg2010}.  For both types 
of peaking background, we determine the relevant proportionality constants 
from Monte Carlo simulations.  We iterate 
our fit to minimize $\chi^2$ and -- at each iteration -- we recalculate 
the internal and external peaking contributions using the $\Bi$ and $\NDDbar$  values obtained in the previous iteration.
These estimated peaking contributions produce yield adjustments of 
${\cal O}(1\%)$.

\begin{table*}[htb]
  \caption{ Single tag efficiencies, yields from data, and peaking
    background estimates for $D\Dbar$ events. The efficiencies include
    the branching fractions for $\piz\to\gamma\gamma$ and
    $\KS\to\pip\pim$ decays. Peaking backgrounds are not included in
    the background shape functions, so the ``Data yield'' values include
    ``Peaking backgrounds''.
    The MC simulations yielded no peaking backgrounds for a few modes,
    indicated by three center dots in the ``Peaking background'' column.
    \label{tab:st-eff-yield}}
\medskip
\begin{ruledtabular}  
\begin{tabular}{lrrc}
Single tag mode  & Efficiency(\%) & Data yield  & Peaking \\
& & & background \\ \hline 
$\Dzkpi$	 & $65.17 \pm 0.11$	 & $75177 \pm 281$	 & $285 \pm 13$\\
$\Dzbarkpi$	 & $65.88 \pm 0.11$	 & $75584 \pm 282$	 & $285 \pm 13$\\
$\Dzkpipiz$	 & $35.28 \pm 0.07$	 & $144710 \pm 439$	 & $296 \pm 17$\\
$\Dzbarkpipiz$	 & $35.62 \pm 0.07$	 & $145798 \pm 441$	 & $296 \pm 17$\\
$\Dzkpipipi$	 & $46.82 \pm 0.09$	 & $114222 \pm 366$	 & $2600 \pm 262$\\
$\Dzbarkpipipi$	 & $47.19 \pm 0.09$	 & $114759 \pm 368$	 & $2600 \pm 262$\\
$\Dpkpipi$	 & $54.92 \pm 0.10$	 & $116545 \pm 354$	 & $\cdots$\\
$\Dmkpipi$	 & $55.17 \pm 0.10$	 & $117831 \pm 356$	 & $\cdots$\\
$\Dpkpipipiz$	 & $28.13 \pm 0.10$	 & $36813 \pm 260$	 & $\cdots$\\
$\Dmkpipipiz$	 & $28.21 \pm 0.10$	 & $37143 \pm 261$	 & $\cdots$\\
$\Dpkspi$	 & $45.63 \pm 0.10$	 & $16844 \pm 137$	 & $81 \pm 22$\\
$\Dmkspi$	 & $45.33 \pm 0.10$	 & $17087 \pm 138$	 & $81 \pm 22$\\
$\Dpkspipiz$	 & $23.95 \pm 0.11$	 & $38329 \pm 262$	 & $110 \pm 52$\\
$\Dmkspipiz$	 & $24.10 \pm 0.11$	 & $38626 \pm 263$	 & $110 \pm 52$\\
$\Dpkspipipi$	 & $32.29 \pm 0.14$	 & $23706 \pm 224$	 & $601 \pm 226$\\
$\Dmkspipipi$	 & $32.60 \pm 0.14$	 & $23909 \pm 225$	 & $601 \pm 226$\\
$\Dpkkpi$	 & $44.52 \pm 0.29$	 & $10115 \pm 123$	 & $\cdots$\\
$\Dmkkpi$	 & $44.66 \pm 0.26$	 & $10066 \pm 123$	 & $\cdots$\\
\end{tabular}
\end{ruledtabular}  
\end{table*}

\section*{Systematic Uncertainties}

We updated systematic uncertainties for the full 818~\pbinv\ data
sample, using methods described in \Ref{cleodhadprd281}. The larger
data sample has led to improvement of some systematic uncertainties
measured in data.  Some other systematic uncertainties were reduced by
improvements in the techniques for their estimation.  The resulting
systematic uncertainties for ST yields for each $\Dz$ and $\Dp$ decay
mode are given in \Tab{syserrors}.

We assign a tracking systematic uncertainty of 0.3\% per $\pipm$ and
0.6\% for $\Kpm$ candidate for all decay modes, including the $\pipm$
produced in $\KS$ decay.  These tracking
uncertainties are correlated among all charged particles.
There is a systematic uncertainty of 0.8\% in the reconstruction
efficiency $\epsilon(\KS)$ for neutral kaons that is correlated among
all $\KS$ candidates.

In further studies following procedures described in Appendix B.5 of
\Ref{cleodhadprd281} we refined our understanding of small differences
between the $\piz$ efficiencies in MC simulations and data.  Based on
these studies, the efficiencies for $\Dzkpipiz$, $\Dpkpipipiz$,  
$\Dpkspipiz$, and their charge conjugates in \Tab{tab:st-eff-yield} 
include a correction factor of
0.939 with uncertainties of 1.3\%, 1.5\%, and 1.3\%, respectively,
reduced from 2\% in \Ref{cleodhadprd281}.

\newcommand{\na}{\multicolumn{1}{c}{\rm ---}}

\begin{table*}
  \caption{Contributions, in percent, to the systematic uncertainties for each ST efficiency-corrected yield, enumerated by decay mode.  The first three modes are $\Dz(\Dzbar)$ and the rest are $\Dp(\Dm)$ modes.  $K$ and $\pi$ are shorthand for the appropriate charged kaons and pions in each decay mode. Each of the uncertainties in the last three rows are not correlated with any other uncertainties. The rest of the uncertainties are fully correlated among all modes within a row, but uncertainties in one row are not correlated with those in another. Efficiency uncertainties (denoted by $\epsilon$) are multiplicative and other (yield) uncertainties are additive.\label{syserrors}}
\begin{ruledtabular}
\begin{tabular}{lddddddddd}

Source          & \multicolumn{1}{c}{~$K\pi$}        & \multicolumn{1}{c}{~$K\pi\piz$}  
                & \multicolumn{1}{c}{~$K\pi\pi\pi$}  & \multicolumn{1}{c}{~$K\pi\pi$}  
                & \multicolumn{1}{c}{~$K\pi\pi\piz$} & \multicolumn{1}{c}{~$\KS\pi$}  
                & \multicolumn{1}{c}{~$\KS\pi\piz$}  & \multicolumn{1}{c}{~$\KS\pi\pi\pi$}  
                & \multicolumn{1}{c}{~$KK\pi$}  \\ \hline
$\epsilon$(Tracking)    & 0.90 & 0.90 & 1.50 & 1.20 & 1.20 & 0.90 & 0.90 & 1.50 & 1.50 \\
$\epsilon(\KS)$         & \na  & \na  & \na  & \na  & \na  & 0.80 & 0.80 & 0.80 & \na  \\
$\epsilon(\piz)$        & \na  & 1.30 & \na  & \na  & 1.50 & \na  & 1.30 & \na  & \na  \\
$\epsilon(\pipm$) PID   & 0.25 & 0.25 & 0.75 & 0.50 & 0.50 & 0.25 & 0.25 & 0.75 & 0.25 \\
$\epsilon(\Kpm$) PID    & 0.30 & 0.30 & 0.30 & 0.30 & 0.30 & \na  & \na  & \na  & 0.60 \\
Lepton veto             & 0.10 & \na  & \na  & \na  & \na  & \na  & \na  & \na  & \na  \\
FSR                     & 0.80 & 0.40 & 0.70 & 0.50 & 0.20 & 0.40 & 0.20 & 0.50 & 0.30 \\
Signal shape            & 0.40 & 0.50 & 0.51 & 0.34 & 0.48 & 0.39 & 0.48 & 0.55 & 0.54 \\
Backg. shape \hfill     & 0.38 & 1.10 & 0.76 & 0.40 & 3.05 & 0.77 & 1.53 & 1.22 & 0.82 \\  
$\DeltaE$ \hfill       & 0.10 & 0.20 & 0.20 & 0.10 & 0.20 & 0.00 & 0.40 & 1.20 & 0.20 \\
Substructure\hfill     & \na  & 0.58 & 1.30 & 0.53 & 0.94 & \na  & 0.42 & 0.62 & 2.60 \\
Mult. cand. \hfill     & 0.00 & 0.70 & 0.00 & 0.00 & 0.20 & 0.20 & 0.00 & 0.00 & 0.00\\

\end{tabular}
\end{ruledtabular}
\end{table*}

Particle identification efficiencies are studied by reconstructing
decays with unambiguous particle content, such as
$\Dz\to\KS\,\pip\pim$ and $\phi\to\Kp\Km$. The decay of $\Dzkpipiz$ is
also used for the study as the $\Km$ and $\pip$ can be distinguished
kinematically.  We require PID for all charged kaons and for all charged pions that are not the daughters of $\KS$ decay.   We utilize the following techniques to account for the small 
differences observed between data and Monte Carlo simulations of PID.
In each final state, we apply an efficiency
correction factor 0.995 per PID-identified $\pipm$ and 0.990 per $\Kpm$.  
We also assign systematic uncertainties of
0.25\% to each PID-identified $\pipm$ and 0.30\% to each $\Kpm$, correlated among all charged PID-identified pions and kaons separately.

We assign a systematic uncertainty of 0.1\% to $\Dzkpi$ single tag yields
to account for the lepton veto requirement.
For FSR we allocate systematic uncertainties of 25\% \cite{fsr} of the correction for each mode, correlated across all modes. 
The systematic uncertainties, (0.4--1.5)\%, for background shapes 
in single tag yields are estimated by using alternative ARGUS parameters.

Other sources of efficiency uncertainty include: the $\Delta E$
requirements (0.0--1.2)\%, for which we examine $\Delta E$ sidebands;
modeling of multiple candidates (0.0--0.7)\%; and
modeling of resonant substructure in multi-body modes (0.4--2.6)\%,
which we assess by comparing simulated momentum spectra to those in
data or changes in ST efficiency due to new measurements of resonant 
substructure.

The effects of quantum correlations between the $\Dz$ and $\Dzbar$
states appear through $\Dz-\Dzbar$ mixing and doubly
Cabibbo-suppressed decays \cite{cpcorr}.  We use the results reported
in Refs.~\cite{ddmix2013} and \cite{coher} to correct the $\Dz$ and 
$\Dzbar$ yields for these effects.  This reduces the systematic uncertainty previously attributed to quantum correlations from 0.8\% to the range (0.1--0.4)\%.

There is no significant deviation from 100\% for the trigger efficiency in the MC simulation of the efficiency, so we no longer assign a systematic uncertainty to it.

The branching fraction fitter~\cite{brfit} takes these systematic 
uncertainties into account, along with ST and DT yields, efficiencies, 
peaking backgrounds, and their statistical uncertainties. We studied the
validity of the fitter and our analysis technique
\cite{cleodhadprd281} using a generic Monte Carlo sample, which had
three times as many events as our data sample.  The results of this
study validated our entire analysis procedure, including the
fitter.

\begin{table*}[htbp]
  \caption{ Results of the fit to our data. The uncertainties quoted
    are statistical and systematic, respectively.  Fractional
    uncertainties are also listed in separate
    columns.\label{tab:fitResultsData}}
\begin{ruledtabular}  
\begin{tabular}{lccc}
Parameter & Fitted value & \multicolumn{2}{c}{Fractional error}\\[-0.6ex] & & Stat.(\%) & Syst.(\%)\\ 
\hline 
$N_{\Dz\Dzbar}$	 & $(2.951 \pm 0.014 \pm 0.035)\times 10^6$	 & $0.5$	 & $1.2$\\
${\cal B}(\Dzkpi)$	 & $(\BDzkpivalue)\%$	 & $0.5$	 & $1.5$\\
${\cal B}(\Dzkpipiz)$	 & $(14.956 \pm 0.074 \pm 0.335)\%$	 & $0.5$	 & $2.2$\\
${\cal B}(\Dzkpipipi)$	 & $(8.287 \pm 0.043 \pm 0.200)\%$	 & $0.5$	 & $2.4$\\
$N_{\Dp\Dm}$	 & $(2.358 \pm 0.014 \pm 0.025)\times 10^6$	 & $0.6$	 & $1.1$\\
${\cal B}(\Dpkpipi)$	 & $(\BDpkpipivalue)\%$	 & $0.6$	 & $1.7$\\
${\cal B}(\Dpkpipipiz)$	 & $(6.142 \pm 0.045 \pm 0.154)\%$	 & $0.7$	 & $2.5$\\
${\cal B}(\Dpkspi)$ 	 & $(1.578 \pm 0.013 \pm 0.025)\%$	 & $0.8$	 & $1.6$\\
${\cal B}(\Dpkspipiz)$	 & $(7.244 \pm 0.053 \pm 0.166)\%$	 & $0.7$	 & $2.3$\\
${\cal B}(\Dpkspipipi)$	 & $(3.051 \pm 0.027 \pm 0.082)\%$	 & $0.9$	 & $2.7$\\
${\cal B}(\Dpkkpi)$	 & $(0.981 \pm 0.010 \pm 0.032)\%$	 & $1.0$	 & $3.2$\\
\end{tabular}
\end{ruledtabular}  
\end{table*}

\section*{Results and Conclusions}

The results of the branching fraction fit are given in Table
\ref{tab:fitResultsData}, where we have listed both statistical and
systematical errors. The correlation matrix for the fitted parameters
is listed in Table \ref{tab:correlationMatrixData}. We also compute
the ratios of branching fractions with respect to the two
``reference'' modes as shown in Table
\ref{tab:fitResultsRatiosData}. The $\chi^2$ of the fit is 46.7 for 52
degrees of freedom.
These results supersede previous CLEO results \cite{cleodhadprl,cleodhadprd281}, obtained utilizing subsets of the full 818~\pbinv\ data sample, and are the most precise results reported to date \cite{pdg2012}.

The $e^+e^-\to D\Dbar$ cross sections are obtained by dividing
$N_{\Dz\Dzbar}$ and $N_{\Dp\Dm}$ by the luminosity of our data set,
$(818.1 \pm 8.2)$ \pbinv. The luminosity was determined using the
procedure described in Appendix C of \Ref{cleodhadprd281}.  We find

\begin{eqnarray}
\sigma( e^+e^-\to \Dz\Dzbar )	 &=& (\sigDzDzbarvalue){~\rm nb}\\
\sigma( e^+e^-\to \Dp \Dm )	 &=& (\sigDpDmvalue){~\rm nb}\\
\sigma( e^+e^-\to D\Dbar )	 &=& (\sigDDbarvalue){~\rm nb}\\
\sigma( e^+e^-\to \Dp \Dm ) / \sigma( e^+e^-\to \Dz\Dzbar )	 &=& \sigDDbarratio
\end{eqnarray}

\noindent where the uncertainties are statistical and systematic,
respectively.  The charged and neutral cross sections have a
correlation coefficient of 0.69 stemming from the systematic uncertainties 
for $\NDzDzbar$, $\NDpDm$, and 
the luminosity measurement.  For this reason, the uncertainty on 
$\sigma( e^+e^-\to D\Dbar )$ is larger than the  
quadratic sum of the charged and neutral cross section uncertainties.

\begin{table*}[htbp]
    \caption{The correlation matrix, including systematic uncertainties,  
     for the fit results for $\NDDbar$ and branching fractions.
     $K$ and $\pi$ are shorthand for the appropriate charged kaons and 
     pions in each decay mode.  
     The parameter order matches that in Table~\ref{tab:fitResultsData}. 
     \label{tab:correlationMatrixData}}
\begin{ruledtabular}
\begin{tabular}{lrrrrrrrrrrr}
\rule[-2ex]{0em}{4ex} & 
$\NDzDzbar$    & $K\pi$ & 
$K\pi\piz$     & $K\pi\pi\pi$ & 
$\NDpDm$       & $K\pi\pi$ & 
$K\pi\pi\piz$  & $\KS\,\pi$ & 
$\KS\,\pi\piz$ & $\KS\,\pi\pi\pi$ &
$KK\pi$ \\ \hline
$\NDzDzbar$	    & $1.00$	 & $-0.56$ & $-0.29$ & $-0.30$  & $0.49$  & $-0.19$  & $-0.11$  & $-0.17$  & $-0.11$  & $-0.08$  & $-0.06$\\ 
$K\pi$           & & $1.00$  & $0.52$  & $0.75$  & $-0.23$  & $0.69$  & $0.45$  & $0.51$  & $0.36$  & $0.51$  & $0.41$\\
$K\pi\piz$       & &    & $1.00$  & $0.43$  & $-0.14$  & $0.41$  & $0.69$& $0.30$  & $0.68$  & $0.31$  & $0.25$\\
$K\pi\pi\pi$     & &    &    & $1.00$  & $-0.13$  & $0.65$  & $0.42$  & $0.47$  & $0.33$  & $0.51$  & $0.37$\\ \hline
$\NDpDm$	         & &    &    &    & $1.00$  & $-0.50$  & $-0.21$  & $-0.51$  & $-0.28$  & $-0.27$  & $-0.24$\\
$K\pi\pi$	     & &    &    &    &    & $1.00$  & $0.50$  & $0.70$  & $0.45$  & $0.63$  & $0.50$\\
$K\pi\pi\piz$    & &    &    &    &    &    & $1.00$  & $0.38$  & $0.65$& $0.37$  & $0.29$\\
$\KS\,\pi$ 	     & &    &    &    &    &    &    & $1.00$  & $0.52$  & $0.63$  & $0.39$\\
$\KS\,\pi\piz$   & &    &    &    &    &    &    &    & $1.00$  & $0.43$& $0.25$\\
$\KS\,\pi\pi\pi$ & &    &    &    &    &    &    &    &    & $1.00$  & $0.35$\\
$KK\pi$          & &    &    &    &    &    &    &    &    &    & $1.00$\\ 
\end{tabular}
\end{ruledtabular}
\end{table*}

\begin{table*}[htbp]
  \caption{ Branching ratios from the fit to our data. The
    uncertainties quoted are statistical and systematic,
    respectively.\label{tab:fitResultsRatiosData}}
\begin{ruledtabular}  
\begin{tabular}{lccc}
Parameter & Fitted value & \multicolumn{2}{c}{Fractional error}\\[-0.6ex] & & Stat.(\%) &
 Syst.(\%)\\ \hline 
${{\calB}(\Dzkpipiz)}/{{\calB}(\Km\pip)}$	 & $3.802 \pm 0.022 \pm 0.073$	 & $0.6$ & $1.9$\\
${{\calB}(\Dzkpipipi)}/{{\calB}(\Km\pip)}$	 & $2.106 \pm 0.013 \pm 0.032$	 & $0.6$ & $1.5$\\
${{\calB}(\Dpkpipipiz)}/{{\calB}(\Km\pip\pip)}$	 & $0.666 \pm 0.006 \pm 0.014$	 & $0.9$ & $2.1$\\
${{\calB}(\Dpkspi)}/{{\calB}(\Km\pip\pip)}$	 & $0.171 \pm 0.002 \pm 0.002$	 & $1.0$ & $0.9$\\
${{\calB}(\Dpkspipiz)}/{{\calB}(\Km\pip\pip)}$	 & $0.785 \pm 0.007 \pm 0.016$	 & $0.9$ & $2.1$\\
${{\calB}(\Dpkspipipi)}/{{\calB}(\Km\pip\pip)}$	 & $0.331 \pm 0.004 \pm 0.006$	 & $1.2$ & $1.8$\\
${{\calB}(\Dpkkpi)}/{{\calB}(\Km\pip\pip)}$	 & $0.106 \pm 0.002 \pm 0.003$	 & $1.4$ & $2.6$\\
\end{tabular}
\end{ruledtabular}  
\end{table*}

For each decay mode $f$ and its charge conjugate $\overline{f}$, 
we obtain the $CP$ asymmetry, 
\Begeqn
A_{CP} (f) \equiv \frac{ \nf - \nfbar }{ \nf + \nfbar } \textrm{,}
\Endeqn
from the single tag yields, $\nf$ and $\nfbar$ obtained after
subtraction of backgrounds and correction for efficiencies
\cite{cleodhadprd281}.  Table~\ref{tab:CPAsymmetries} gives the values
of $A_{CP}(f)$ obtained from the full 818~\pbinv\ data sample.  No
mode shows evidence of $CP$ violation at the level of the
uncertainties, which are of order 1\% for all modes.  Standard Model
estimates of $CP$ violation are at most a few tenths of a percent
\cite{Bianco:2003vb} and we are not sensitive to asymmetries at this
level.

\begin{table*}
\caption{$CP$ asymmetry for each decay mode, in percent.\label{tab:CPAsymmetries}}
\begin{ruledtabular}  
\begin{tabular}{lr}
Mode	 & $CP$ Asymmetry (\%)\\ \hline 
$\Dzkpi$	 & $0.3 \pm 0.3 \pm 0.6$\\
$\Dzkpipiz$	 & $0.1 \pm 0.3 \pm 0.4$\\
$\Dzkpipipi$	 & $0.2 \pm 0.3 \pm 0.4$\\
$\Dpkpipi$	 & $-0.3 \pm 0.2 \pm 0.4$\\
$\Dpkpipipiz$	 & $-0.3 \pm 0.6 \pm 0.4$\\
$\Dpkspi$	 & $-1.1 \pm 0.6 \pm 0.2$\\
$\Dpkspipiz$	 & $-0.1 \pm 0.7 \pm 0.2$\\
$\Dpkspipipi$	 & $0.0 \pm 1.2 \pm 0.3$\\
$\Dpkkpi$	 & $-0.1 \pm 0.9 \pm 0.4$\\
\end{tabular}
\end{ruledtabular}  
\end{table*}

In summary, we report measurements of three $\Dz$ and six $\Dp$
branching fractions and the production cross sections
$\sigma(\Dz\Dzbar )$, $\sigma(\Dp\Dm)$, and $\sigma(D\Dbar)$ using a
sample of 818 ${\rm pb}^{-1}$ of $e^+e^-\to D\Dbar$ data obtained at
$\Ecm=3774 \pm 1$ MeV.

\section*{Acknowledgments}

We gratefully acknowledge the effort of the CESR staff in providing us
with excellent luminosity and running conditions.  This work was
supported by the National Science Foundation, the U.S.\ Department of
Energy, the Natural Sciences and Engineering Research Council of
Canada, and the U.K Science and Technology Facilities Council. X. Shi
thanks the High Energy Group National Taiwan University National
Science Council where part of this work was completed.

\end{document}